\documentclass[letterpaper,10pt,conference,twocolumn]{IEEEtran}
\IEEEoverridecommandlockouts

\usepackage{ifthen}

\usepackage[tbtags]{amsmath} 
\usepackage{amssymb}  
\usepackage{verbatim} 
\usepackage{amsxtra}  

\usepackage{enumerate}

\usepackage{amsfonts}
\usepackage{color}

\usepackage{subcaption}
\usepackage{wasysym}

\usepackage{cite}

\usepackage{mathtools}
\mathtoolsset{showonlyrefs=true}

\newcommand{\isi}[1]{}
\newcommand{\isiincl}[2]{}

\newcommand{\googleincl}[2]{}
\newcommand{\googleinclabs}[3]{}

\usepackage{tikz}
\usetikzlibrary{shapes,arrows,decorations.markings,positioning}

\tikzstyle{plant} = [draw, fill=red!5, rectangle, 
    minimum height=3em, minimum width=6em]
\tikzstyle{block} = [draw, fill=blue!5, rectangle, 
    minimum height=3em, minimum width=6em]
\tikzstyle{sum} = [draw, fill=yellow!10, circle, node distance=1cm]
\tikzstyle{coord} = [coordinate]
\tikzstyle{gain} = [draw, fill=red!5, regular polygon, regular polygon sides=3, shape border rotate=-90]
\tikzstyle{pinstyle} = [pin edge={to-,thick,black}]

\tikzstyle{BitPipe} = [thick, decoration={markings,mark=at position
   1 with {\arrow[semithick]{open triangle 60}}},
   double distance=1.4pt, shorten >= 5.5pt,
   preaction = {decorate},
   postaction = {draw,line width=1.4pt, white,shorten >= 4.5pt}]

\usepackage{epsfig, graphicx, psfrag}

\usepackage[mathcal]{euscript}
\usepackage[cal=boondox]{mathalfa}

\usepackage{amsthm}

\theoremstyle{plain}
\newtheorem{thm}{Theorem}
\newtheorem{lemma}{Lemma}

\theoremstyle{definition}

\newtheorem*{defn*}{Definition}
\newtheorem{scheme}{Scheme}
\newtheorem*{scheme*}{Scheme}

\theoremstyle{remark}
\newtheorem{remark}{Remark}

\providecommand{\secref}[1]{Sec.~\ref{#1}}

\providecommand{\remref}[1]{Rem.~\ref{#1}}
\providecommand{\figref}[1]{Fig.~\ref{#1}}

\providecommand{\schemeref}[1]{Sch.~\ref{#1}}

\newcommand{\ie}{i.e.}
\newcommand{\eg}{e.g.}

\newcommand{\reals}{\mathbb{R}}

\DeclareMathOperator{\R}{Re}

\newcommand{\bm}[1]{\mbox{\boldmath{$#1$}}}

\newcommand{\SNR}{\mathrm{SNR}}
\newcommand{\SDR}{\mathrm{SDR}}

\newcommand{\Comment}[1]{}
\newcommand{\old}[1]{}
\newcommand{\rem}[1]{}

\newcommand{\hX}{\hat{X}}
\newcommand{\heta}{\hat{\eta}}

\newcommand{\tX}{\tilde X}

\newcommand{\tY}{\tilde Y}

\newcommand{\bS}{\text{\bm S}}

\newcommand{\hbX}{\hat{\bX}}

\newcommand{\bC}{\text{\bf C}}

\newcommand{\bI}{\text{\bf I}}

\newcommand{\bX}{\boldsymbol{X}}

\newcommand{\bQ}{\mathbf{Q}}

\providecommand{\bB}{\mathbf{B}}
\providecommand{\bA}{\mathbf{A}}

\newcommand{\Norm}[1]{\left\| #1 \right\|}

\providecommand{\comment}[1]{}

\providecommand{\norm}[1]{\Norm{#1}}
\providecommand{\fmod}[1]{f_{\alpha,\beta,\Delta}{\left(#1\right)}}
\providecommand{\fmodnoarg}{f_{\alpha,\beta,\Delta}}

\newcommand{\beqn}[1]{\begin{eqnarray}\label{#1}}
\newcommand{\eeqn}{\end{eqnarray}}
\newcommand{\beq}[1]{\begin{equation}\label{#1}}
\newcommand{\eeq}{\end{equation}}

\makeatletter
\newcommand{\vast}{\bBigg@{4}}
\newcommand{\Vast}{\bBigg@{5}}
\makeatother

\providecommand{\E}[1]{\mathbb{E} \left[ #1 \right]}
\providecommand{\CE}[2]{\mathbb{E} \left[ #1 \middle| #2 \right]}

\providecommand{\Eig}{\lambda}

\providecommand{\SNRb}{\SNR_\mathrm{B}}
\providecommand{\costX}{q}
\providecommand{\costXvec}{\bQ}
\providecommand{\costU}{{r}}

\providecommand{\lVec}{\boldsymbol{L}}

\providecommand{\XtVec}{\tilde{\boldsymbol{X}}}
\providecommand{\XVec}{\bX}
\providecommand{\hXVec}{\hat{\bX}}

\providecommand{\htX}{\hat{\tX}}
\providecommand{\oJ}{\bar{J}}

\usepackage{ifthen}
\usepackage{comment}

\usepackage{mathtools}
\mathtoolsset{showonlyrefs=true}

\usepackage{comment}

\newcommand{\VersionLength}{short}

\providecommand{\ver}{\ifthenelse{\equal{\VersionLength}{long}}}
\providecommand{\nver}{\ifthenelse{\equal{\VersionLength}{short}}}
\providecommand{\first}[2]{#1}

\newtheorem{theorem}{Theorem}

\first{}{
	\setlength{\abovedisplayskip}{3.5pt}
	\setlength{\belowdisplayskip}{4pt}

	\renewcommand{\baselinestretch}{0.98}
}

\ver{\usepackage{subfig}}{}
\ver{\usepackage{hyperref}}{\usepackage{url}}

\begin{document}
\title{\LARGE \bf Schemes for LQG Control over Gaussian Channels with \\ Side Information}
	
\author{Omri Lev and Anatoly Khina%
    \thanks{This work has received funding from the European Union's Horizon 2020 research and innovation programme under the Marie Sk\l odowska-Curie grant agreement No 708932.
    The work of O.\ Lev was supported in part by the Yitzhak and Chaya Weinstein Research Institute  for Signal Processing. 
    \newline {\color{black} } The authors are with the Department of Electrical Engineering---Systems, Tel Aviv University, Tel Aviv, Israel 6997801. \mbox{E-mails}: \mbox{{\tt \{omrilev@mail,anatolyk@eng\}.tau.ac.il}}}
}
\maketitle
	

\begin{abstract}
	We consider the problem of controlling an unstable scalar linear plant over a power-constrained additive white Gaussian noise (AWGN) channel, where the controller/receiver has access to an additional noisy measurement of the state of the control system.
	To that end, we view the noisy measurement as side information and 
	recast the problem to that of joint source--channel coding with 
	side information at the receiver. 
	We argue that judicious modulo-based schemes improve over their linear counterparts and allow to avoid a large increase in the transmit power due to the ignorance of the side information at the sensor/transmitter.
	We demonstrate the usefulness of our technique for the settings where i) the sensor is oblivious of the control objectives, control actions and previous controller state estimates, ii) the system output tracks a desired reference signal that is available only at the controller via integral control action. 
\end{abstract}

\ver{
}{}

\allowdisplaybreaks

\section{Introduction}
\label{s:intro}

Recent advances in wireless communications have brought us to the verge of the era of the Internet of Things, 
which raises, in turn, the demand for new and improved techniques for control of cyberphysical systems over noisy communication media
\cite{FranceschettiMinero:ElementsNCS,NetworkedControlSurvey_ProcIEEE,BansalBasar:JSCC4Control,SilvaDerpichOstergaard:ECDQ4Control,SchenatoSinopoliFranceschettiPoolaSSS,Nair:L1,TatikondaSahaiMitter,YukselBasarBook,LQGoverAWGN:linear:NoBraslavsky,StavrouSkoglund:ControlWZ:TechRep2019,KostinaHassibi:RDF4Control:AC,JSCC4Control:AC2019}.
In contrast to traditional control, in which the system components (sensor, plant, and controller) are colocated, 
the components of CPS may be non-colocated and communicate instead over noisy channels.

In this work, we consider the setting where the controller observes the system state corrupted by noise via an internal sensor, while it also receives descriptions of the observations of the state from an external sensor over an additive white Gaussian noise (AWGN) channel. We concentrate on a simple fully-observable discrete-time linear quadratic Gaussian (LQG) control setting.

To exploit the internal measurements of the controller, we view them as side information (SI) that is known at the controller (that also acts as a receiver) but not at the (external) sensor (that also acts as a transmitter). This interpretation allows us to appeal to to zero-delay joint source--channel coding (JSCC) techniques. Specifically, we build on techniques that utilize modular arithmetic \cite{JointWZ-WDP,ZamirBookKochmanChapter,Tuncel_ZeroDelayJSCCwithWZ}, 
which allow to mimic the operation of two-sided SI schemes despite not knowing the SI at the transmitter, and avoid most of the power increase (alternatively, distortion increase) due to the lack of SI knowledge  at the sensor.
We incorporate these techniques into the schemes previously developed for LQG control over AWGN channels without SI \cite{BansalBasar:JSCC4Control,LQGoverAWGN:linear:NoBraslavsky,JSCC4Control:AC2019}, 
and show an improvement over linear techniques, recently suggested by Stavrou and Skoglund \cite{StavrouSkoglund:ControlWZ:TechRep2019}, thus prove that linear techniques are suboptimal in the presence of controller SI.

In many practical scenarios, the sensor is oblivious of past control actions, controller state estimates and control-objectives---the linear quadratic regulator (LQR) weights. By viewing these signals as additional SI that is known to the controller and the sensor, we show that our proposed technique readily applies to this scenario as we as well.

We then extend our treatment to the setting where the controller aims the system state to track a desired reference trajectory (unknown at the sensor) instead of driving the former to zero (as per simple LQG control) via integral action \cite[Ch.~6.4]{AastromMurray:book}. To that end, we recast this problem, again, as that of control with controller SI, where the reference trajectory takes the role of SI.

The rest of the paper is organized as follows. We present the notation used in this work in \secref{ss:notation} and formulate the problem of interest in \secref{s:model}. Schemes for low-delay JSCC with SI are detailed in \secref{s: Low Delay JSCC}, and are subsequently subsequently used in \secref{s:control design} to develop control policy with controller SI.
We then use these technique to develop a scheme that for the setting of a sensor that is oblivious of the control actions, controller state estimates and control objectives (LQR weights), in \secref{s: Control with Cost Uncertainty}. We further extend the technique to work for the setting where the controller aims the system state to track a desired reference trajectory (unknown at the sensor) in \secref{s: LQG Reference Tracking}. The exposition of the proposed technique and schemes in Secs.\ \ref{s:control design}--\ref{s: LQG Reference Tracking} is supplemented by simulations that demonstrate the improvement of the modular-arithmetic schemed over their linear counterparts.
We conclude the paper with a discussion about extensions to vector systems and channels in \secref{s:Discussion}.
	
\vspace{-.1\baselineskip}
\subsection{Notations}
\label{ss:notation}
\vspace{-.15\baselineskip}

Throughout the paper, $\norm{\cdot}$ denotes the Euclidean norm.
We denote temporal sequences by $a_{1:t} \triangleq \left( a_1, \ldots, a_t \right)$. 
Random values are denoted by capital letters. 
We denote $a \| b \triangleq ab / (a + b)$, and 
$[\cdot]_{\Delta}$ denotes the the modulo-$\Delta$ operation, \ie, $[x]_{\Delta} = x - \Delta\cdot\mathrm{round}\left(x/\Delta\right)$.
We use the notation $\fmod{X}$ for the class of modulo-based encoding functions 
$\fmod{X} \triangleq \alpha (X - [X]_{\Delta}) + \beta [X]_{\Delta}$ for 
parameters $\alpha, \beta, \Delta$. 
	

	
\vspace{-.05\baselineskip}
	\section{Problem Statement}
	\label{s:model}
\vspace{-.15\baselineskip}
	
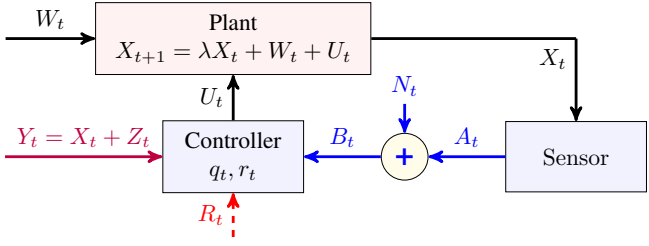
\begin{figure}
	    \centering
    	{\center
    	\newcommand{\sensor}{\mathrm{Sensor}}
    	\newcommand{\state}{X_t}
    	\newcommand{\observeSig}{Y_t}
    	\newcommand{\driveNoise}{W_t}
    	\newcommand{\channelNoise}{\color{blue} N_t}
    	\newcommand{\channelIn}{\color{blue} A_t}
    	\newcommand{\channelOut}{\color{blue} B_t}
    	\newcommand{\contAction}{U_t}
    	\newcommand{\observeCoeff}{\bC}
    	\newcommand{\plant}{X_{t+1} = \Eig \state + \driveNoise + \contAction}
        \newcommand{\channelConstraint}{}
        \newcommand{\objective}{\costX_t, \costU_t}
        \newcommand{\channelNoiseFB}{\color{red} \nu_i}
        \newcommand{\channelInFB}{\color{red} \alpha_i}
        \newcommand{\channelOutFB}{\color{red} \beta_i}
    	\resizebox{\columnwidth}{!}{\begin{tikzpicture}[auto, arrow/.style={very thick, ->, >=stealth'},node distance=.2\columnwidth,>=latex']
        \node [coord] (input) {};
        \node [plant, right of = input, node distance=.385\columnwidth] (plant) 
        {$\begin{array}{c} 
        	  \text{Plant} 
	       \\ \plant 
	      \end{array}$
	    };
        \node [coord, right of=plant, node distance = .58\columnwidth] (sum) {};

        \node [block, below of=sum, node distance = .2\columnwidth] (enc) {$\sensor$};
        \node [block, below of=plant] (dec) 
        {$\begin{array}{c} 
        	  \text{Controller} 
	       \\ \objective 
	      \end{array}$
	    };
        \node [sum, left of=enc, node distance = .29\columnwidth] (channel) {\bf \Large \color{blue} +};

        \draw[arrow] (sum) -- node [left, pos=.25] (yArrow) {$\state$} (enc);
        \node [above of=channel, node distance = .12\columnwidth] (channelNoise) {$\channelNoise$};
        \draw[arrow,color=blue] (channelNoise) -- node {} (channel);
        \draw[very thick,-] (plant) -- node {} (sum);
        
        \draw[arrow] (input) -- node {$\driveNoise$} (plant);
        \draw[arrow] (dec) -- node {$\contAction$} (plant);
        \draw[arrow,color=blue] (enc) -- node [above] {$\channelIn$} node [below] {\footnotesize $\channelConstraint$} (channel);
        \draw[arrow,color=blue] (channel) -- node [above] (outarrow) {$\channelOut$} (dec);

        \node [left of=dec, node distance = .4\columnwidth] (SI_source) {\bf \Large \color{purple}};   
        \draw[arrow, color=purple] (SI_source) -- node {$Y_t = X_{t}+Z_{t}$} (dec);    
	   
        \node [below of=dec, node distance = .15\columnwidth] (Ref_Input) {\bf \Large \color{red}};   
        \draw[arrow, color=red, dashed] (Ref_Input) -- node {$R_t$} (dec);   
        

    \end{tikzpicture}
    }}
	\vspace{-1.4\baselineskip}
		\caption{Scalar discrete-time linear plant controlled over an AWGN channel with SI that is available at the controller. $R_t$ is a reference trajectory, set by the controller.}
    \label{fig:Control_over_AWGN}
\vspace{-1\baselineskip}
	\end{figure}

	The control--communications setting treated in this work is depicted in \figref{fig:Control_over_AWGN}. 
	
	We consider a discrete-time scalar linear plant dynamics 
\vspace{-.3\baselineskip}
	\begin{align}
	\label{eq:x_t:recursion}
	    X_{t+1} &= \Eig X_t + W_t + U_t \,,
	& t = 0, \ldots, T \,,
	\end{align}
	across a finite time horizon $T$, 
	where $X_t \in \reals$ is the (scalar) state at time $t$, $W_{t} \in \reals$ is an additive white Gaussian noise (AWGN) (system disturbanceז) of power $\sigma_W^2$, $\Eig$ is the (known) scalar open-loop gain, and $U_{t} \in \reals$ is the control action applied at time $t$. 
    We assume zero initial state conditions: $X_0 = 0$. 

	In contrast to traditional control settings, 
	the sensor is not colocated with the controller and transmits to it over AWGN 
	\begin{align}
	\label{eq:channel:forward}
	    &B_t = A_t + N_t 
	\end{align}
	per each control sample,
	where $B_t$ is the channel output, $A_t$ is the channel input subject to a power constraint $\E{A_t^2} \leq 1$, and $N_t$ is an AWGN of power $1/\SNR$, where $\SNR > 0$ is the channel signal-to-noise ratio (SNR).
	


	We further assume a SI signal that is available to the controller, but not to the sensor.
	This signal might be available, for example, when the controller observes the system state via an internal sensor.
	The SI is assumed to be a noisy version of the current source sample $X_t$, and is given by 
	\begin{align}
	\label{eq:External SI model}
	    Y_t = X_t + Z_t ,
	\end{align}	
	where 
	$Z_t$ is an AWGN of power $\sigma_Z^2$ 
	independent of $\{(X_t, N_t)\}$.

\begin{remark}[Stabilizability]
    Since the external SI measurements \eqref{eq:External SI model} constitutes a noisy observation of the state variable $X_t$, the system is stabilizable based on this measurement alone [without any transmission over the channel \eqref{eq:channel:forward}].
    Thus, the system is stable for any $\SNR \geq 0$ in \eqref{eq:channel:forward}, 
    in contrast to the setting without SI where the $\SNR$ has to be high enough for the system to be stabilizable (see, \eg, \cite{JSCC4Control:AC2019}).
\end{remark}
\begin{remark}[Two-sided SI]
     The scenario where the SI $Y_t$ is known to both the sensor and the controller (two-sided SI) is equivalent to the case without SI of e.g.\ \cite{JSCC4Control:AC2019,LQGoverAWGN:linear:NoBraslavsky,KostinaHassibi:RDF4Control:AC}, w.r.t.\ to a (Gaussian) source that is equal to $X_t$ given $Y_{1:t}$.
\end{remark}

	As in traditional LQG control, 
	we wish to minimize the following average-stage control cost: 
	\begin{align}
	    \Bar{J}_T = &
     \frac{1}{T} \E{\costX_{T+1} X_{T+1} + \sum_{t=1}^T \left(\costX_t X_t^2 + \costU_t U_t^2 \right)} , 
	\label{eq:control cost}
	\end{align}
	for some non-negative control weights $\{\costX_t\}, \{\costU_t\}$. 

It will be further instructive to consider the fixed-weights steady-state regime: $\costX_t \equiv \costX$, $\costU_t \equiv \costU, T \to \infty$. We denote by 
\begin{align}
\label{eq:cost:SS}
    \oJ_\infty \triangleq \lim_{T \to \infty} \oJ_T 
\end{align}
the steady-state average-stage control cost, for this setting.
    
To that end, we next review known results on low-delay joint source--channel coding (JSCC) with SI.
	
	
	\section{Zero-Delay JSCC with SI}
	\label{s: Low Delay JSCC}
	
	In this section, we review known results 
	for transmitting a zero-mean Gaussian source sample $X$ of power $P_X$ 
	over a single AWGN channel use \eqref{eq:channel:forward}, 
	where the receiver is equipped with Gaussian SI 
	\begin{align}
	    \label{eq:JSCC_SI}
	    	Y = X + Z
	\end{align}
	that is available at the receiver but not at the transmitter \cite[Ch.~11]{ElGamalKimBook}, where $Z$ is Gaussian independent of $X$ of power $P_Z$. 
	
	The goal of the transmitter is to convey the source sample
	$X$ to the receiver with minimal average quadratic distortion 
	\begin{align}
	\label{eq:distortion:def}
	    D \triangleq \E{(X - \hX)^2}, 
    \end{align}
    where $\hX$ is the estimate of the receiver given the channel output $B$ and the SI $Y$.
    
    Since the distortion is proportional to the signal power $P_X$, a popular metric to compare between different JSCC schemes is the signal-to-distortion ratio ($\SDR$) which is defined as 
    \begin{align}
        \label{eq:SDR:def}
        \SDR \triangleq \frac{P_X}{D} \,.
    \end{align}
%

\subsection{Without SI}

Without side information \eqref{eq:JSCC_SI}, the optimal distortion \eqref{eq:distortion:def} of transmitting a single Gaussian source sample over a single AWGN channel use is given as follows.

\begin{theorem}
\label{thm:RDF_without_SI}
    The minimal distortion $D^*_\mathrm{noSI}$ \eqref{eq:distortion:def} in conveying a single Gaussian source sample of average power $P_X$ over a single AWGN channel use with SNR $\SNR$ is equal to 
    \begin{align}
    \label{eq:Rate Distortion AWGN}
        D^*_\mathrm{noSI} = \frac{P_X}{1 + \SNR} ,
    \end{align}	 
    and the corresponding SDR is 
        $\SDR^{*}_{\mathrm{noSI}} 
        = 1 + \SNR$.
\end{theorem}
The converse part of this proof is a straightforward consequence of the source--channel separation principle \cite[Ch~3.9]{ElGamalKimBook},
whereas the direct is established by transmitting the source as is over the channel up to a power adjustment \cite{Elias57:JSCC:BW-expansion}.


\subsection{With Two-Sided SI}

The setting where the SI is available at both the transmitter and the receiver 
is a simple adaptation of the no-SI setting, 
as it is equivalent to conveying a Gaussian source sample without SI of power 
$Var(X|Y) = P_X \| P_Z$,
as the latter is the minimum mean square error
power of the (Gaussian) estimation error of $X$ given $Y$.
\begin{lemma}
    The minimal distortion $D^*_\mathrm{both}$ \eqref{eq:distortion:def} in conveying a single Gaussian source sample of average power $P_X$ over a single AWGN channel use with SNR $\SNR$, 
    where the receiver has access to a side information \eqref{eq:JSCC_SI} 
    is equal to 
    \begin{align}
    \label{eq:Rate Distortion Two Sided AWGN}
        D^*_\mathrm{both}(P_X,P_Z) = \frac{P_X \| P_Z}{1 + \SNR}, 
    \end{align}
    and the corresponding $\SDR$ is given by
    \begin{align}
    \label{eq:Rate Distortion Two Sided AWGN SDR}
        \SDR^*_\mathrm{both}(P_X,P_Z,\SNR) 
        = \left( 1 + \frac{P_X}{P_Z} \right) \cdot (1 + \SNR).
    \end{align}

\end{lemma}

\subsection{With Receiver SI}
\label{ss:JSCC:SI@Rx}

We now consider the setting where the SI \eqref{eq:JSCC_SI} is available to the receiver but not to the transmitter. 

Clearly, the distortion $D^*_\mathrm{Rx}$ of this setting is bounded between 
those with two-sided SI and without SI:
\begin{align}
\label{eq:distortion:bounds}
    D^*_\mathrm{both} \leq D^*_\mathrm{Rx} \leq D^*_\mathrm{noSI} \,,
\end{align}
which is equivalent to 
    $\SDR^{*}_{\mathrm{noSI}} \leq \SDR^{*}_{\mathrm{Rx}} \leq \SDR^{*}_{\mathrm{both}}$.

Moreover, in the infinite-delay setting where multiple i.i.d.\ Gaussian source samples are processes and transmitted over multiple AWGN channel uses, the lower bound in \eqref{eq:distortion:bounds} is attained. 
However, when restricted to our (causal) case of interest, this lower bound is unattainable \cite{LevKhina:CRDF:ISIT2020,WeissmanElGamal_FiniteLookAhead}, 
although the exact value of $D^*_\mathrm{Rx}$ remains unknown.


In the rest of the section, we presents schemes for the zero-delay JSCC with receiver SI setting.


    
\begin{figure}[t]
	\vspace{-.5\baselineskip}
	\includegraphics[width=\columnwidth]{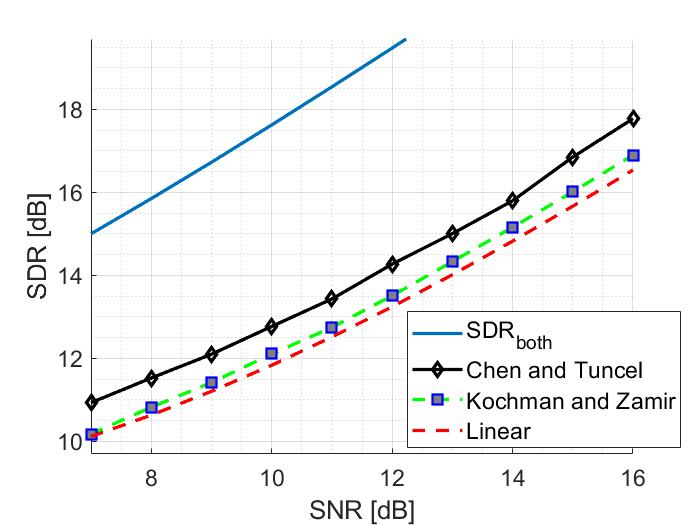}
	\centering
	\vspace{-1\baselineskip}
	\caption{$\SDR$ vs $\SNR$ for the model in \secref{s: Low Delay JSCC} with $\sigma_Z^2 = 1/9$.
	``Chen--Tuncel" and ``Kochman--Zamir" are the schemes from \cite{Tuncel_ZeroDelayJSCCwithWZ,JointWZ-WDP} with parameters $\left(\Delta,\alpha,\beta\right) = (3.15,0.8,-1.15),(2.15,1.05,0)$, respectively.
	}
	\label{fig: Gaussian with SI}
	\vspace{-.1\baselineskip}
\end{figure}

We first present a simple analog transmission.

\begin{scheme}[Linear]
\label{scheme:JSCC:linear}
\ 

    \textit{Transmitter:} 
   Sends 
	        $\displaystyle A = \frac{1}{\sqrt{P_X}} X $. 
	
	\textit{Receiver:}
	Upon receiving the channel output $B = A + N$ and the SI signal $Y = X + Z$, calculates the resulting minimum mean square error (MMSE) estimator 
	
	\begin{align}
    \label{eq:Scalar Analog, MSE}
    \\[-2\baselineskip]
        \hX = \frac{1}{P_Z + \frac{P_X + P_Z}{\SNR}} \left( \sqrt{P_X}P_Z B + \frac{P_X}{\SNR} Y \right) , 
	\end{align}
    whose SDR is equal to 
        $\displaystyle \SDR_{\mathrm{Lin}} = \frac{P_X}{D_\mathrm{Lin}}$,
    where
    \begin{align}
    \label{eq:D:lin}
        D_\mathrm{Lin} &= 
        \frac{P_X}{1 + \SNR + P_X/P_Z}
        = \left. P_X \middle\| P_Z \middle\| \frac{P_X}{\SNR} \right. 
    \end{align}
\end{scheme}
Clearly, $D_\mathrm{Lin}$ of \eqref{eq:D:lin} is strictly higher than $D^*_\mathrm{Both}$ of \eqref{eq:Rate Distortion Two Sided AWGN}.
%
%
Moreover, linear schemes in the presence of receiver SI are known to be sub-optimal
\cite{JointWZ-WDP,ZamirBookKochmanChapter,Tuncel_ZeroDelayJSCCwithWZ}.
We describe next modulo-based JSCC schemes with SI.

\begin{scheme}[Modulo-based]
\label{scheme:JSCC with non-linear functions}
\ 

	\textit{Transmitter:} 
	 Sends 
	\begin{align}
	    \label{eq:Kochman Zamir + Tuncel Encoder}
	    A &= \alpha \left(\Bar{X} - \left[\Bar{X}\right]_{\Delta}\right) + \beta \left[\Bar{X}\right]_{\Delta} \triangleq \fmod{\Bar{X}}\,,
	\end{align} 
	where $\Bar{X} \triangleq \frac{X}{\sqrt{P_X}}$ and $\alpha$, $\beta$ and $\Delta$ are chosen such that $\E{A^2} \leq 1$.

	\textit{Decoder:}
	Upon receiving the channel output $B$ and the SI $Y$,
	evaluates the MMSE estimate:
	    $\hX = \CE{X}{B, Y}$.
\end{scheme}

Note that 
    \schemeref{scheme:JSCC with non-linear functions}
    subsumes the schemes suggested by Kochman and Zamir \cite{JointWZ-WDP}, 
    and by Chen and Tuncel \cite{Tuncel_ZeroDelayJSCCwithWZ} for different choices of $\alpha$,  $\beta$, and $\Delta$, who, addition to the optimal (MMSE) decoder, suggested also suboptimal decoders that are more amenable to analysis.
    Furthermore, by choosing $\Delta \to \infty$, $\alpha = P_X^{-1/2}$, and $\beta = 0$, \schemeref{scheme:JSCC with non-linear functions} reduces to \schemeref{scheme:JSCC:linear}.

Further note that, for every specific choice of $\alpha,\beta, \Delta$ and decoder $\hX(B,Y)$, the scheme yields different $\SDR$ values which, in general, are not amenable to an analytical calculation but can be calculated numerically instead. We denote the distortion and the SDR of \schemeref{scheme:JSCC with non-linear functions}, by (resp.)
\begin{align}
\label{eq:def:SDR:modulo}
    \!\!\! D_{\alpha,\beta,\Delta}(P_X,P_Z,\SNR) 
    \text{\ \ and\ \ }
    \SDR_{\alpha,\beta,\Delta}(P_X,P_Z,\SNR). \quad
\end{align}


Indeed, as is evident from \figref{fig: Gaussian with SI}, 
the schemes of Chen and Tuncel \cite{Tuncel_ZeroDelayJSCCwithWZ}, and Kochman and Zamir \cite{JointWZ-WDP} outperform the linear scheme.

\section{Control Policies with Controller SI}
\label{s:control design}

In this section, we construct control policies that aim to minimize the control cost \eqref{eq:control cost}, 
by relying on the schemes of \secref{s: Low Delay JSCC} and the traditional (without noisy channels) LQG control setup 
\cite{BertsekasControlVol1}. We assume in this section, that the sensor knows the LQR weights $\{ \costX_t, \costU_t \}$
and is made aware of $\hX^r_{t-1}$ at time $t$ (and can therefore construct $U_{t-1}$ and $\hX^r_{t|t-1}$) for the construction of $A_t$. \textit{We part with these assumptions in \secref{s: Control with Cost Uncertainty}}.

We start by presenting a simple linear scheme that achieves the optimum for the case where the SI $Y_t$ is available both at the sensor and the controller via a simple adaptation of the scheme of \cite{BertsekasControlVol1,LQGoverAWGN:linear:NoBraslavsky,JSCC4Control:AC2019}, in \secref{s:full control access}.
We then construct schemes for the setting where only the controller has access to the SI $Y_t$ by employing the schemes of \secref{ss:JSCC:SI@Rx}.

\subsection{With Two-sided SI}
\label{s:full control access}

Here we assume that the SI $Y_{1:t}$ is known also at the sensor,
and design a simple linear scheme that is optimal in this scenario. 
\textit{This scheme will be used as a benchmark to test the performance of other schemes (with lesser sensor SI).}

The scheme and its optimality are simple adaptations of the setting without SI \cite{LQGoverAWGN:linear:NoBraslavsky,JSCC4Control:AC2019}:
In the presence of two-sided SI, the scheme is equivalent to the setting without SI with respect to 
source process $X_t$ given the side information process $Y_t$.
Thus, by looking at the state innovations given $Y_t$ the scheme is equivalent to the setting without SI.
    
\begin{scheme}[Linear scheme with Two-sided SI]
\label{sch:control:two-sided}
\
    
    \textit{Sensor:} At time $t$: 
	\begin{itemize}
	\item 
	    Calculates the prediction error (innovation) $\tX_{t|t-1}$ of the controller given the SI $Y_t$ and the prediction error $\hX^r_{t|t-1}$ before receiving $Y_t = \hX^r_{t|t-1} + \tX^r_{t|t-1} + Z_t$: 
         \begin{align}
             \label{eq:Observer, Full Access}
             \tX_{t|t-1} &= X_t - \E{X_t|Y_{1:t}, \hX^{r}_{t|t-1}}
          \\ &= \tX^r_{t|t-1} - \rho_t \left( Y_t - \hX^r_{t|t-1} \right)
          \\ &= (1 - \rho_t) \tX^r_{t|t-1} - \rho_t Z_t \,,
         \end{align}
        whose average power is 
        \begin{align}
            \label{eq:ConstantDef}
            P_{t|t-1} &\triangleq \E {\left(\tX_{t|t-1}\right)^2} = \left. \sigma_Z^2 \middle\| P^r_{t|t-1} \right. , 
        \end{align}        
        and where
        \begin{align}
            \rho_t &\triangleq \frac{P^{r}_{t|t-1}}{P^{r}_{t|t-1} + \sigma_{Z}^{2}} = \frac{P_{t|t-1}}{ \sigma_Z^2 }
        \end{align}
        is the correlation coefficient between $Y_t$ and $\tX^r_{t|t-1}$.

    \item 
        Transmits the prediction error $\tX_{t|t-1}$ given the SI with appropriate power adjustment:
         \begin{align}
             A_t &= \frac{\tX_{t|t-1}}{\sqrt{P_{t|t-1}}} \:.
         \end{align}
	\end{itemize}
		
	
	
	
	\textit{Controller:} At time $t$:
	\begin{itemize}
	\item 
		Estimates the prediction error $\tX_{t|t-1}$ via MMSE estimation given 
		\begin{align}
		    B_t = A_t + N_t = \frac{\tX_{t|t-1}}{\sqrt{P_{t|t-1}}} + N_t
		\end{align}
		as follows.
		\begin{align}
		    \htX_{t|t-1} = \frac{\sqrt{P_{t|t-1}}}{1 + 1/\SNR} B_t .
		\end{align}
	\item 
		Constructs an estimate $\hX^r_{t|t}$
		given $\htX_{t|t-1}$, the SI 
		$Y_t$ and $\hX^r_{t|t-1}$ (which are a sufficient statistic of $(Y_{1:t},B_{1:t})$):
		\begin{align}
		    \label{eq:Controller, Full Access, state estimation}
		    &\hX^{r}_{t|t} = \hX^{r}_{t|t-1} + \rho_t \left( Y_t - \hX^r_{t|t-1} \right) + \htX_{t|t-1} , 
		\end{align}
		and calculates its MMSE:
		\begin{align}
		    \label{eq:Controller, Full Access, estimation error}
	    	& P^{r}_{t|t} = \frac{\sigma_Z^2 \| P^r_{t|t-1}}{1+\SNR} = \frac{P_{t|t-1}}{1+\SNR} \:. 
		\end{align} 
	\item 
		Generates the control action, the state-prediction at time $(t+1)$ and its MMSE according to
		\begin{subequations}
		\label{eq:controller:signal-generate}
		\noeqref{eq:controller:signal-generate:U,eq:controller:signal-generate:hX^r,eq:controller:signal-generate:P^r}
		\begin{align}
		    U_t &= -L_t\hX^{r}_{t|t},
		\label{eq:controller:signal-generate:U}
		 \\ \hX^{r}_{t+1|t} &= \Eig \hX^{r}_{t|t} + U_t,
		\label{eq:controller:signal-generate:hX^r}
		 \\ P^r_{t+1|t} &= \Eig^2 P^{r}_{t|t} + \sigma_W^2 \,,
		\label{eq:controller:signal-generate:P^r}
		\end{align}  
		\end{subequations}
		
	    where the control gain $L_{t}$ is given by the Riccati recursion (see, e.g. \cite{BertsekasControlVol1}) with boundary condition $s_T = \costX_{T+1}$: 
    	\begin{align}
	    \label{eq:Riccati}
    	    L_t &= \frac{\Eig s_{t+1}}{s_{t+1}+\costU_{t+1}} \,,
    	  & s_t &= \frac{\Eig^2 \costU_{t+1} s_{t+1}}{s_{t+1} + \costU_{t+1}} + \costX_{t+1} \,. \quad
    	\end{align}
	\end{itemize}
\end{scheme}

\subsection{With Receiver SI}
\label{s:zero delay JSCC codes}

Here we assume that the SI $Y_t$ at time $t$ is available at the controller but not at the sensor.
We first introduce a na\"ive linear scheme, followed by an improved modulo-based scheme.



\begin{scheme}[Linear-based]
\label{sch:control:linear}
\

   \textit{Sensor}: At time $t$,
    \begin{itemize}
    \item
        Calculates the prediction error of the receiver, $\tX^r_{t|t-1} \triangleq X_t - \hX^r_{t|t-1}$, and its power, $P^r_{t|t-1} = \E{\left(\tX^r_{t|t-1}\right)^2}$.
    \item 
        Transmits the prediction error with appropriate power adjustment:
	       $
	        A_t = \frac{1}{\sqrt{P^r_{t|t-1}}} \tX^r_{t|t-1} \,.
	       $
    \end{itemize}
    
	\textit{Controller}: At time $t$: 
	\begin{itemize}
    \item 
        Estimates the prediction error $\tX^r_{t|t-1}$ via MMSE estimation\footnote{Since we employ only linear operations at the sensor, all variables are jointly Gaussian and hence all the MMSE estimators are linear.} given
        \begin{align}
            B_t = A_t + N_t 
            = \frac{1}{\sqrt{P^r_{t|t-1}}} \tX^r_{t|t-1} + N_t ,
        \end{align}
        and the SI $\tY_t = Y_t - \hX^r_{t|t-1} = \tX^r_{t|t-1} + Z_t$,
        as in  
        \eqref{eq:Scalar Analog, MSE}:
        \begin{align}
            \htX^r_{t|t-1} =  \frac{1}{\sigma_Z^2 + \frac{P^r_{t|t-1} + \sigma_Z^2}{\SNR}} \left[ \sqrt{P^r_{t|t-1}}\sigma_Z^2 B_t + \frac{P^r_{t|t-1}}{\SNR} \tY_t \right]
        \nonumber
        \end{align}

    \item
	    Updates the state estimate and its mean square error (MSE):
	    \begin{subequations}
	    \label{eq:@Rx-linear}
	    \noeqref{eq:@Rx-linear:power}
	    \begin{align}
	        \hX^{r}_{t|t} &= \hX^{r}_{t|t-1} + \htX_{t|t-1},
	    \label{eq:@Rx-linear:estimate}
	     \\ P^{r}_{t|t} &=  \left. P^r_{t|t-1} \middle\| \sigma_Z^2 \middle\| \frac{P^r_{t|t-1}}{\SNR} \right. 
	     = \frac{\left(1-\rho_t\right)P^r_{t|t-1}}{1+(1-\rho_t)\SNR} \qquad
	    \label{eq:@Rx-linear:power}
	    \end{align} 
	    \end{subequations}
	    where \eqref{eq:@Rx-linear:power} is according to \eqref{eq:D:lin} and $\rho_t$ is given in \eqref{eq:ConstantDef}; it is evident from \eqref{eq:@Rx-linear:estimate} that $\htX^r_{t|t-1} = \tX^r_{t|t-1} - \tX^r_{t|t}$.
    \item
	    Generates the control action and the next-state prediction according to \eqref{eq:controller:signal-generate}.
	\end{itemize}
\end{scheme}
	
	

	
We now improve the linear scheme by employing \schemeref{scheme:JSCC with non-linear functions} instead of \schemeref{scheme:JSCC:linear}.
	
\begin{scheme}[modulo-based]
\label{sch:control:modulo}
\

    \textit{Sensor:} At time $t$:
    \begin{itemize}
    \item 
        Calculates the prediction error of the receiver, $\tX^r_{t|t-1} \triangleq X_t - \hX^r_{t|t-1}$, and its power, $P^r_{t|t-1} = \E{\left(\tX^r_{t|t-1}\right)^2}$. 
    \item 
        Applies $\fmodnoarg$ of \eqref{eq:Kochman Zamir + Tuncel Encoder} to the prediction error after normalizing its power
        \begin{align}
            \label{eq:Observer, JSCC}
            A_t &= \fmod{ \frac{ \tX^r_{t|t-1} }{ \sqrt{P^r_{t|t-1} } }} , 
        \end{align} 
        with coefficients $\left(\Delta,\alpha,\beta\right)$ that satisfy the power constraint $\E{A_t^2} \leq 1$ (to be determined in the sequel).
   \end{itemize}
	   
	\textit{Controller}: At time $t$:
	\begin{itemize}
    \item 
	    Constructs a state estimate given the SI and channel-output histories, $Y_{1:t}$ and $B_{1:t}$, respectively:
	    \begin{align}
	        \hX^{r}_{t|t} &= \E{X_t|B_{1:t},Y_{1:t}} = \hX^{r}_{t|t-1} + \E{\tX^r_{t|t-1}|B_{1:t},\tY_{1:t}}
	   \nonumber
	     \\&= \hX^{r}_{t|t-1} + \E{\tX^r_{t|t-1}|B_t,\tY_t}.
	    \label{eq:@Rx-JSCC:Markov}
	    \end{align} 
        
        Where \eqref{eq:@Rx-JSCC:Markov} holds as long as the estimation errors and the encoder output forms a Markov chain, and thus is true whenever the assumption on perfect feedback of the estimations $\hX^r_{t|t-1}$ holds.
         
    \item 
	    Updates the state estimate MSE according to the SDR of the underlying JSCC scheme \schemeref{scheme:JSCC with non-linear functions} [recall \remref{rem:SDRvsSNR}, \eqref{eq:def:SDR:modulo}]
	    \begin{align}
	        \label{eq:ControllerJSCC:MSE}
	        P^{r}_{t|t} = \frac{P^{r}_{t|t-1}}{\SDR_{\alpha,\beta,\Delta}(P^{r}_{t|t-1},P_Z,\SNR)}.
	    \end{align}
	    
        
    \item
	    Generates the control action and the next-state prediction according to \eqref{eq:controller:signal-generate}.
	\end{itemize}
\end{scheme}

We are left with determining the coefficients $\alpha, \beta, \Delta$. 
As the prediction error power 
$P^r_{t|t-1}$ changes across time (during its initial transient response, until it converges to steady-state operation), one may use time-varying coefficients $\alpha_t,\beta_t,\Delta_t$ by optimizing them with respect to the instantaneous correlation coefficient $\rho_t$. 

However, since the system converges to steady-state operation, we use time-constant coefficients instead, in \schemeref{sch:control:modulo}, that are optimized for steady-state operation and satisfy the average power constraint $\frac{1}{T}\sum_{t=1}^{T}\E{A_t^2} \leq 1$.


	
\begin{remark}
    \label{rem:SubOptimalKlaman}
    In contrast to the linear schemes (Schs.\ \ref{sch:control:two-sided} and \ref{sch:control:linear}) where all the signals and the estimation errors are jointly Gaussian, the estimation errors in the (non-linear) modulo-based scheme (\schemeref{sch:control:modulo}) are not Gaussian. Nonetheless, numerical investigation suggests that they are nearly Gaussian and the design for Gaussian variables works essentially the same as if they were Gaussian. A similar observation was reported in the rate-mismatched setting in \cite{JSCC4Control:AC2019}.
\end{remark}
     
	

\subsection{Fixed-Coefficients Steady-State Operation}
\label{s:control:SS}

For the case of constant cost coefficients $\costX_t \equiv \costX, \costU_t \equiv \costU$ the next theorems holds. Their proofs are essentially the same as those for the setting without SI \cite{KostinaHassibi:RDF4Control:AC,JSCC4Control:AC2019} and are therefore omitted in the interest of space.
\begin{thm}[Achievability]
\label{thm:AchievableCost}
    The infinite-horizon average-stage control cost $\oJ_\infty$ \eqref{eq:cost:SS} of the system of model \secref{s:model} is bounded from above by
    \begin{align}
        \label{eq: infinite horizon cost}
        \oJ_\infty \leq S\sigma_W^2 + \frac{Q + (\Eig^2-1)S}{\SDR_{\infty} - \Eig^2} \sigma_W^2 ,
    \end{align}
    where $S$ is the steady state solution of the Riccati equation \eqref{eq:Riccati}, and $\SDR_{\infty}$ is the SDR of the underlying JSCC scheme in steady-state.
\end{thm}

\begin{thm}[Impossibility]
\label{thm:LowerBoundCost}
The optimal achievable infinite-horizon average-stage LQG cost of the scalar control system \secref{s:model} is
bounded from below by 
    \begin{align}
        \oJ_\infty \geq S\sigma_W^2 + \frac{Q + (\Eig^2-1)S}{\SDR^\mathrm{both}_{\infty} - \Eig^2} \sigma_W^2 ,
    \end{align}
    where 
	\begin{align}
	    \label{eq:SDR_forBound_TwoSided}
	    \SDR^{\mathrm{both}}_{\infty} = \lim_{t \to \infty} \frac{P^r_{t|t-1}}{D_{\mathrm{both}}(P^r_{t|t-1},\sigma_Z^2)} = \frac{1 + \SNR}{1 - \rho_{\infty}} 
	\end{align}
	and $\rho_{\infty} = \lim_{t \to \infty} \rho_t$.
\end{thm}


\begin{figure}[t]
	\vspace{-.5\baselineskip}
	\includegraphics[width=\columnwidth]{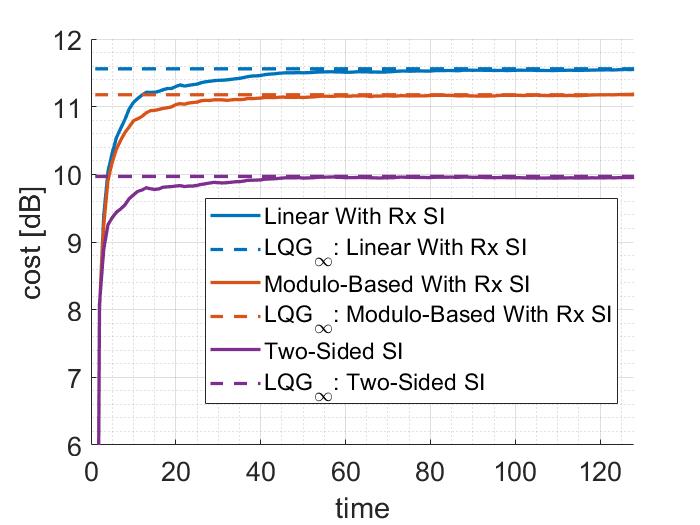}
	\centering
	\vspace{-1.5\baselineskip}
	\caption{Control cost evolution across time for Schs.\ \ref{sch:control:two-sided}--\ref{sch:control:modulo} for 
	$\SNR = 6 \text{dB}, \sigma_W^2 = 1, \sigma_Z^2 = 1/8, \Eig = 2$, $Q = 5$, $R = 1$. The parameters used in \schemeref{sch:control:modulo} for transmission are $\alpha = 0.75, \beta = -1.18, \Delta = 3.2$. 
    The plots are averaged over $2^{10}$ runs.}
	\label{fig: Control, no Feedback, constant cost}
	\vspace{-.38\baselineskip}
\end{figure}

\begin{remark}
\label{rem:SDRvsSNR}
    We use the more common (biased) variant of the SDR definition throughout this work, in contrast to the correlation-unbiased estimator (CUBE) SDR, $\SDR_\mathrm{CUBE}$, that was preferred in \cite{JSCC4Control:AC2019}, the relation between the two being 
    $\SDR = 1 + \SDR_\mathrm{CUBE}$.
\end{remark}

\subsection{Simulations}
\label{s:Simulations without feedback}
    
We simulate a system with $\Eig = 2$ and $\sigma^2_W = 1$.
We further use $Q = 5, R = 1$ (all parameters are assumed to be known at both nodes). The noise in the receiver SI channel \eqref{eq:External SI model} is taken to be with power $\sigma_Z^2 = 1/8$. 
The performance of Schs.\ \ref{sch:control:two-sided}--\ref{sch:control:modulo} for these parameters are presented in \figref{fig: Control, no Feedback, constant cost}, where the optimized parameters that were selected in \schemeref{sch:control:modulo} are 
$\alpha = 0.75, \beta = -1.18, \Delta = 3.2$.
This figure clearly demonstrates a performance boost due to the introduction of the modulo-based component.

    


\section{Control without Control-Objectives or Control-Actions Knowledge at the Sensor}
\label{s: Control with Cost Uncertainty}

In the previous section, we have assumed that the weights $\{\costX_t\}$ and $\{\costU_t\}$ are known both to the sensor and the controller, and that the state estimates and control actions of the controller are sent back via a perfect feedback to the sensor for the generation of the next channel input.

However, in many practical scenarios, the sensor is unlikely to known the control objectives $\{\costX_t\}$ and $\{\costU_t\}$ (up to maybe a region to which these may belong to) that are determined by the controller according to the system requirements and might be changed during operation, or have perfect (continues-amplitude) feedback of the control action and/or state estimate of the controller. 

In this section, we construct schemes that are oblivious to the control objectives of the controller (albeit know to what region they may belong to) and lack any feedback of the control actions and/or state estimates of the controller. To that end, we view the control actions and controller state estimates as additional SI in Schs.\ \ref{sch:control:linear} and \ref{sch:control:modulo} of \secref{s:control design}.
        
\subsection{Problem Statement}
\label{s: Control with Cost Uncertainty - problem statement}

We assume that the sensor is not aware of the control objectives, \ie, $\{\costX_t\}$ and $\{\costU_t\}$, 
but knows that the resulting (optimal) LQR coefficient $L_t$ of \eqref{eq:Riccati} is in the region $[L^{\text{min}}_{t},L^{\text{max}}_{t}]$. The sensor, further has no access to $u_t$ or $\hX_{t|t-1}$.

The aim of the sensor is to transmit to the controller in a way that will be robust to this lack of knowledge.
To that end, we reinterpret the lack of knowledge as additional SI that is known at the controller but not at the sensor, as follows.
%
\begin{subequations}
\noeqref{eq:state-side information:regular,eq:state-side information:estimates,eq:state-side information:SI}
\label{eq:state-side information}
\begin{align}
    X_t &= \Eig X_{t-1} + W_{t} + U_{t}
\label{eq:state-side information:regular}
 \\ &=\Eig(X_{t-1} - \hX^{r}_{t-1|t-1}) + W_t + (\Eig - L_t)\hX^{r}_{t-1|t-1} \qquad
\label{eq:state-side information:estimates}
 \\ &=\tX^r_{t|t-1} + \underbrace{(\Eig - L_t)\hX^{r}_{t-1|t-1}}_{\mathrm{SI}}.
\label{eq:state-side information:SI}
\end{align} 
\end{subequations}
%
%
With the state power breaking down as 
\begin{subequations}
\label{eq:Recursive state power}
\begin{align}
    P_{X_t} &= P^r_{t|t-1} + (\lambda - L_t)^2 P_{\hX_{t-1|t-1}}
\label{eq:Recursive state power:basic}
  \\ &= \sigma^2_W + (\Eig - L_t)^2 P_{X_{t-1}} + L_t(2\Eig-L_t) P^r_{t-1|t-1} \quad\:\:
\label{eq:Recursive state power:extended}
\end{align} 
\end{subequations}
where $P_{X_t}$ and $P_{\hX_{t-1|t-1}}$ are the powers of $X_t$ and $\hX_{t-1|t-1}$, respectively, 
\eqref{eq:Recursive state power:basic} is due to \eqref{eq:state-side information:SI} and the orthogonality~principal, 
and \eqref{eq:Recursive state power:extended} is due to \eqref{eq:controller:signal-generate:P^r} and the orthogonality~principal.


\subsection{Control Design}
\label{s: Control with Cost Uncertainty - control design}	
    
To circumvent the the uncertainty in the control cost due to the lack of knowledge of the SI in \eqref{eq:state-side information}, we consider two setups for the choice of the transmit power over the channel: 
\begin{itemize}
\item 
    \textit{Worst-case scaling:} Under this conservative setup, the sensor works with respect to the value $L_t \in [L_t^{\min}, L_t^{\max}]$  that induces the largest transmit power, namely, as if \mbox{$L_t = L^{\min}_t$}.
        
\item 
    \textit{Randomized scaling:} Under this regime, a probability distribution over $L_t$ is assumed. Here, for simplicity of exposition, we assume that $L_t$ is uniformly distributed over its uncertainty interval $[L_t^{\min}, L_t^{\max}]$.
\end{itemize}

        The controller designs its estimate based on the selected setup, by calculating the transmit power according to \eqref{eq:Recursive state power}.
	
	The suggested control schemes are extensions 
	of Schs.\ \ref{sch:control:linear}--\ref{sch:control:modulo}.

	\begin{scheme}[Control under cost uncertainty]
	\label{scheme:Control_CostUncertain}
	
	\
	
    \textit{Sensor:} At time $t$:
	\begin{itemize}
	\item
		Calculates the current state power $P_{X_t}$ according to \eqref{eq:Recursive state power} with the scaling chosen with respect to the setup used.
	\item 
		Applies $\fmodnoarg$ of \eqref{eq:Kochman Zamir + Tuncel Encoder} to the state variable after normalizing its power:
	    \begin{align}
	      \label{eq:Observer, Zero Access, cost uncertainty}
	      A_t &= \fmod{\frac{X_t}{\sqrt{P_{X_t}}}}
        \\  &= \fmod{\frac{\tX^{r}_{t|t-1} + (\Eig - L_t)\hX^{r}_{t|t-1}}{\sqrt{P_{X_t}}}} .
	    \end{align}	
	\end{itemize}

	\textit{Controller}: At time $t$:
	\begin{itemize}
	\item 
    Constructs a state estimate given the external SI $Y_t$, the state prediction $\hX^{r}_{t|t-1}$ and the channel output~$B_t$:
    \begin{align}
        \label{eq:@Rx-JSCC_Uncertain:StateEstimation}
        \hX^{r}_{t|t} &= \hX^{r}_{t|t-1} + \E{\tX^r_{t|t-1}|B_t,\tY_t,\hX^{r}_{t|t-1}} .
    \end{align} 
 \item
	 Updates the state estimate MSE according to \eqref{eq:ControllerJSCC:MSE}.
 \item
	 Generates the control action and the next-state prediction according to \eqref{eq:controller:signal-generate}.
	\end{itemize}
\end{scheme}
	
		    Note that when $\Delta \to \infty$, $\alpha = P_X^{-1/2}$ the function $\fmodnoarg$ turns to simple linear encoder and \eqref{eq:@Rx-JSCC_Uncertain:StateEstimation} turns to the MMSE estimator \eqref{eq:Scalar Analog, MSE}.
	
	\begin{remark}
	\label{rem:KalmanSubOptimal_Uncertain}
	Since we do not assume perfect feedback from the controller to the sensor, the encoder output is not a Markov chain and thus \eqref{eq:@Rx-JSCC_Uncertain:StateEstimation} is the true MMSE estimate only when the encoder is linear. Furthermore, following \remref{rem:SubOptimalKlaman}, the estimation errors in the (non-linear) modulo-based scheme are not Gaussian. Nonetheless, numerical investigation suggests that they are nearly Gaussian and the design for Gaussian variables works essentially the same as if they were Gaussian. 
	    
	\end{remark}
\subsection{Simulations}
\label{s: Control with Cost Uncertainty - simulations}

\begin{figure}[t]
	\vspace{-.5\baselineskip}
	    \centering
	    \includegraphics[width=\columnwidth]{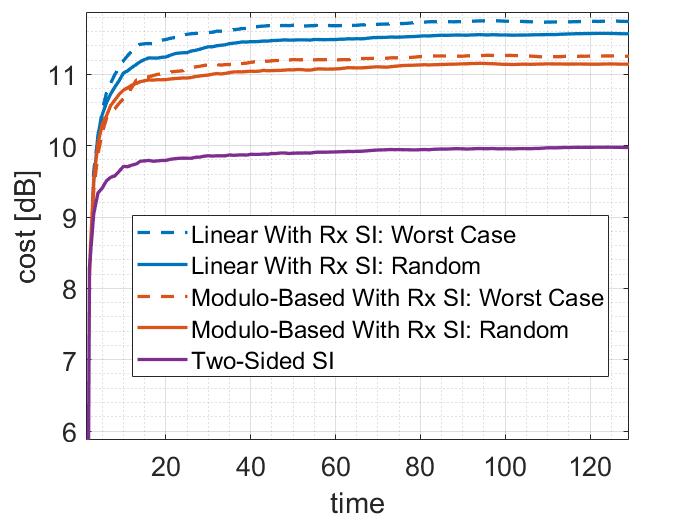}
	    \vspace{-1\baselineskip}
	\caption{Control cost evolution across time for Sch.\ \ref{scheme:Control_CostUncertain} and its linear counterpart for 
	$\SNR = 6 \text{dB}, \sigma_W^2 = 1, \sigma_Z^2 = 1/6, \Eig = 2$, $Q = 5$, $R = 1, L_t^{\min} = L_t/3, L_t^{\max} = 3L_t$. The parameters used in the modulo-based scheme for transmission are $\alpha = 0.75, \beta = -1.18, \Delta = 3.2$. The performances of both schemes is simulated under both scaling setups and are compared to the two-sided scheme (\schemeref{sch:control:two-sided}) with perfect control-objectives knowledge at the sensor.
    The plots are averaged over $2^{10}$ runs.}
\label{fig: Control uncertainty}
\end{figure}

As we have seen in \secref{s:Simulations without feedback}, 
the modulo-based scheme enjoys better performance compared to its linear counterpart.
Under the current model, in which the SI is stronger, the performance boost promised by the modulo-based scheme is expected to be even greater. Indeed, by carrying simulations for the same parameters as in \secref{s:Simulations without feedback}, we observe in \figref{fig: Control uncertainty} a greater gap between the modulo- and linear-based schemes; we use $L^{\min}_t = L_t/3$ and $L^{\max}_t = 3 L_t$, where $L_t$ is the true~value.

   

\section{Extension: LQG Control with Integral Control Action and Unknown Reference Input}
\label{s: LQG Reference Tracking}

In this section, we extend our treatment to the setting where the controller aims the system state $X_t$  to track a reference signal that is known (or determined) by the controller but is unknown at the sensor---a common setup in practice.

To that end, we appeal to the traditional LQG control with integral action \cite[Ch.~6.4]{AastromMurray:book}, 
in which the system state is augmented by the integral of the reference-tracking error signal;\footnote{The integration is analogous to the integral component in PID controllers.} 
this method allows tracking a general reference input, without calibrating the controller to a predefined reference~level.

Since the reference signal is known only at the controller but unlikely to be known at the sensor, 
we recast this signal as additional SI. 

    
\subsection{Problem Statement}
\label{s: LQG Reference Tracking - Problem Description}

We consider the setup of \secref{s: Control with Cost Uncertainty}, namely, the model of \secref{s:model}, where the sensor is not aware of the control objectives $\{\costX_t\}, \{\costU_t\}$, nor of the control action and controller state estimate signals $Y_t$ and $(\lambda - L_t)\hX_{t-1|t-1}$, respectively.
However, instead of driving the state $X_t$ to zero, the controller wishes $X_t$ to follow a reference trajectory $R_t$ that may change across time. Since the reference $R_t$ is time-varying, one \textit{cannot} simply replace the term $\E{\costX_t X_t^2}$ in the control cost \eqref{eq:control cost} with $\E{\costX_t (X_t - R_t)^2}$, as the design should be universal (robust) with respect to the values of the signal $R_t$.
    
Following the traditional LQG control framework for this setting \cite[Ch.~6.4]{AastromMurray:book}, 
we augment the state space \eqref{eq:x_t:recursion} by the accumulated sum of tracking error signal, 
$\eta_t$:
 \begin{align}
\label{eq:Tracking Error}
    \eta_t &= \eta_{t-1} + R_{t-1} - X_{t-1}, & \eta_0 &= R_0 .
 \end{align}	

 The augmented state space model is given by 
\begin{subequations}
\label{eq:Augmented State Space}
\noeqref{eq:Augmented State Space:recursion,eq:Augmented State Space:values}
\begin{align}
    \bX_{t+1} &= \bA \bX_t + \bB U_t + \begin{pmatrix}W_t \\ 0\end{pmatrix} + \begin{pmatrix}0 \\ R_t\end{pmatrix}
%
\label{eq:Augmented State Space:recursion}
    \\
    \bX_t &= \begin{pmatrix}X_t \\ \eta_t\end{pmatrix}, \quad
    \bA = \begin{pmatrix} \Eig&0 \\ -1&1\end{pmatrix},  \quad 
    \bB = \begin{pmatrix}1 \\ 0\end{pmatrix}.
\label{eq:Augmented State Space:values}
\end{align}
\end{subequations}
For the construction of 
$U_t$,
\eqref{eq:controller:signal-generate:U} is replaced with
\begin{align}
    \label{eq:RefTracking_ControlAction}
    U_t = -\lVec_t^{T}\hXVec^r_{t|t}.
\end{align}
where $\hXVec^r_{t|t} \triangleq \begin{pmatrix}\hX^{r}_{t|t}&\heta^{r}_{t|tk}\end{pmatrix}^T$ is the estimated state vector. 
The augmented LQG cost that we wish to minimize 
is given by 
\begin{align}
\label{eq:LQG Vector}
    \oJ_T = \frac{1}{T} \E{\bX_T^T\costXvec_{T+1}\bX_T + \sum_{t=1}^T\left(\bX_t^T\costXvec_t\bX_t+\costU_t U_t^2\right)} \quad
\end{align}
where $\costXvec_t$ is a positive semidefinite weight matrix, and $\costU_t \geq 0$.
    

\subsection{Control Design}
	
	The main difference with respect to the SI schemes presented in \secref{s: Control with Cost Uncertainty} is the vector form of the problem, and the calculation of the state power $P_{X_t}$. 
	Throughout the derivation we use the notations: 
    \begin{align}
        \label{eq:cov definition}
        \XtVec^{r}_{t|t} &\triangleq \XVec_t - \hXVec^{r}_{t|t} \triangleq \begin{pmatrix} \tX^{r}_{t|t} \\ \tilde{\eta}^{r}_{t|t}\end{pmatrix},
        C_{\bX_t} \triangleq \E{\XVec_t\XVec_t^T},\\
        C^{r}_{t|t} &\triangleq \E{\left(\XtVec^{r}_{t|t}\right)\left(\XtVec^{r}_{t|t}\right)^T} \triangleq \begin{pmatrix}P^{r}_{t|t} && 0 \\ 0 && P^{r,\eta}_{t|t}\end{pmatrix} ,
    \end{align}
    where $C^{r}_{t|t}$ is diagonal since the tracking-error estimate at time $t$ depends on the state estimates up to time $t-1$, and thus is uncorrelated with the state estimate at time $t$, according to \cite[Lem.~3.2]{TatikondaSahaiMitter}.
	By substituting \eqref{eq:RefTracking_ControlAction} into \eqref{eq:Augmented State Space}, we get:
    \begin{align}
    \label{eq:Augmented state for power}
	     \XVec_{t+1} = \bA\XVec_t - \bB\lVec_t^{T}(\XVec_t - \XtVec^{r}_{t|t}) + \begin{pmatrix}W_t \\ 0\end{pmatrix} + \begin{pmatrix}0 \\ R_t\end{pmatrix}
    \end{align}
    The state mean is, therefore,\footnote{Recall that $R_t$ is a signal determined by the controller and is therefore not random but rather deterministic but unknown to the sensor.}
\begin{align}
\label{eq:Expectation Reusrive}
     \E{\XVec_{t+1}} = \left(\bA - \bB\lVec_t^T\right)\E{\XVec_{t}} 
     + \begin{pmatrix}0 \\ R_t\end{pmatrix} 
\end{align}
and the state covariance is equal to 
\begin{align}
\nonumber
     &\E{\left\{\XVec_{t+1} - \begin{pmatrix}0 \\R_t\end{pmatrix}\right\}\left\{\XVec_{t+1} - \begin{pmatrix}0 \\R_t\end{pmatrix}\right\}^{T}} - \begin{pmatrix} \sigma_W^2 && 0 \\ 0 && 0\end{pmatrix}\\
\label{eq:Covariance Reusrive}
     &+
     \left(\bA - \bB\lVec_t^T\right) C_{\bX_t} \left(\bA - \bB\lVec_t^{T}\right)^T  + \bB\lVec_t^TC^{r}_{t|t}\lVec_t\bB^T \\
     &+\left(\bA - \bB\lVec_t^T\right)C^{r}_{t|t}\lVec_t \bB^T + \bB\lVec_t^TC^{r}_{t|t}\left(\bA - \bB\lVec_t^T\right)^T .
\end{align}
By rearranging \eqref{eq:Expectation Reusrive} and \eqref{eq:Covariance Reusrive}, we get a recursive description of the state covariance $C_t$ and consequently also a recursive formula for $P_{X_t} = \E{X_t^2}$.
As the control objectives and the reference $R_t$ are unknown to the sensor, we may work with the same scaling setups of \secref{s: Control with Cost Uncertainty - control design} also with respect to the reference trajectory $R_t$ which can be assumed to belong to $[R^{\min}_t, R^{\max}_t]$.

    

\begin{figure}[t]
		\vspace{-.5\baselineskip}
		\includegraphics[width=\columnwidth]{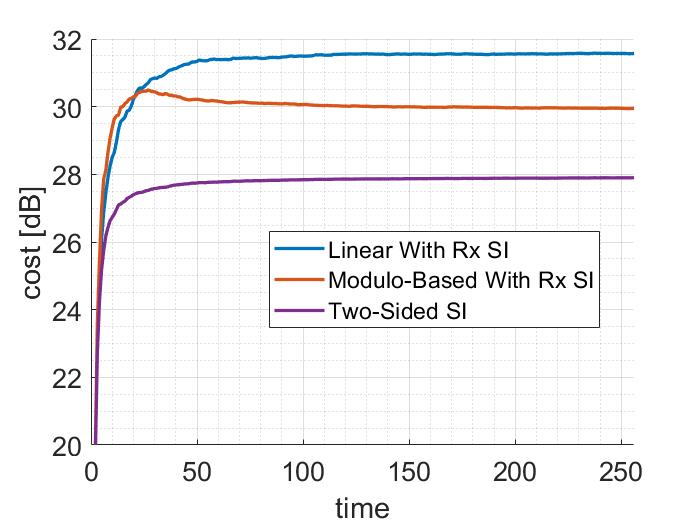}
		\centering
		\vspace{-1.5\baselineskip}
	\caption{Control cost evolution across time for Sch.\ \ref{scheme:Control_ReferenceTracking} and its linear counterpart for 
	$\SNR = 9 \text{dB}, \sigma_W^2 = 1, \sigma_Z^2 = 12, \Eig = 2$, $\costXvec_t= 5 \bI_2, \costU_t = 1, R_t = 10, L_t^{\min} = L_t/3, L_t^{\max} = 3L_t, R_t^{\min} = R_t/3, R_t^{\max} = 3R_t$. The parameters used in the modulo-based scheme are $\alpha = 0.75, \beta = -1.18, \Delta = 3.2$. The performances of both schemes is simulated under both scaling setups and are compared to the two-sided scheme (\schemeref{sch:control:two-sided}) with perfect control-objectives and reference knowledge at the sensor.
    The plots are averaged over $2^{10}$ runs.}
		\label{fig: reference tracking uncertainty, randomized}
		\vspace{-.38\baselineskip}
\end{figure}
	
    \begin{scheme}[Control with Integral Action]
    \label{scheme:Control_ReferenceTracking}
	
	\
	
    \textit{Sensor:} At time $t$:
	\begin{itemize}
    \item
	    Calculates the current state power $P_{X_t}$ according to \eqref{eq:Covariance Reusrive} and \eqref{eq:Expectation Reusrive} with the scaling chosen with respect to the setup used.
    \item 
	    Applies $\fmodnoarg$ of \eqref{eq:Kochman Zamir + Tuncel Encoder} to the state variable after normalizing its power:
	    \begin{align}
	    \label{eq:Observer, Zero Access, referenceTracking}
	    A_t &= \fmod{\frac{X_t}{\sqrt{P_{X_t}}}} .
	    \end{align}	
	\end{itemize}
	\textit{Controller:} At time $t$:
	    	
	    	\begin{itemize}
	    	    \item 
                    Constructs a state estimate given the external SI $Y_t$, the state predictions $\hbX^{r}_{t|t-1}$ and the channel output~$B_t$:
		            \begin{align}
		                \label{eq:Controller, Integral Action}
		                \hX^{r}_{t|t} = \hX^{r}_{t|t-1} + \E{\tX^r_{t|t-1}|B_t,\tY_t,\hbX^{r}_{t|t-1}}.
		            \end{align} 

		        \item
                    Predicts the next tracking error (with $\hat{\eta}_{0} = R_0$): 
                    \begin{align}
                    \label{eq:Controller, Integral Action, RxSI tracking error}
	                    \hat{\eta}_{t+1} &= \hat{\eta}_{t} + (R_t - \hX_{t|t}),  
                    \end{align}
                \item
                    Updates the state-estimation MSE $P^{r}_{t|t}$ according to \eqref{eq:ControllerJSCC:MSE}, and the tracking-error MSE according to
                    \begin{align}
                        \label{eq:Controller_RefInput_TrackErrorUpdate}
                        P^{r,\eta}_{t|t} = P^{r,\eta}_{t-1|t-1} + P^{r}_{t-1|t-1}.
                    \end{align}
		        
		        \item
		            Computes the control action according to the standard vector LQR recursion \cite{BertsekasControlVol1}: 
                    \begin{align}
                        \label{eq:Riccati matrix}
                        \bS_t &= \bA^{T}\bS_{t+1}\bA -         \frac{(\bA^T\bS_{t+1}\bB)(\bB^T\bS_{t+1}\bA)}{\costU_t+\bB^T\bS_{t+1}\bB} +         \costXvec_t, 
                    \nonumber
                     \\ \lVec_{t+1} &= \frac{\bB^T\bS_t\bA}{\bB^T\bS_t\bB + \costU_t}, \:\:
                        U_t = -\lVec^T_t\hXVec^{r}_{t|t}, \:\:
                        \bS_T = \costXvec_{T+1}.
                    \nonumber
	                \end{align}
		        
		        \item 
	                Predicts the next state according to
	                \eqref{eq:controller:signal-generate:hX^r} and \eqref{eq:controller:signal-generate:P^r}.
	    	\end{itemize}
	\end{scheme}
	
	\subsection{Simulations}
	\label{s:LQG Tracking Simulations}
	Simulation of the last scheme, for $R_t = 10, \costXvec_t = 5 \bI_2$, $\costU_t = 1$, and uncertainty regions $[\frac{\lVec_t}{3},3\lVec_t], [\frac{R_t}{3},3R_t]$, is given in \figref{fig: reference tracking uncertainty, randomized}.
	For the sake of presentation we plot only the state part of the cost. We see that the modulo-based schemes shows improvement over their linear counterparts.
\section{Future Research}
\label{s:Discussion}

Modulo-based schemes are widely used in information~theory in multi-user multi-input multi-output (MIMO) communication setups, where inter-channel and inter-source interference effects are reduced by treating them as SI that is known at the transmitter or receiver\cite{ZamirBook}. Extending the schemes presented in this work for MIMO systems and channels seems, therefore, plausible and is currently under investigation.
    
	
	
	
	
	
    \bibliographystyle{IEEEtran}
	\bibliography{toly}	

\begin{thebibliography}{10}
\providecommand{\url}[1]{#1}
\csname url@samestyle\endcsname
\providecommand{\newblock}{\relax}
\providecommand{\bibinfo}[2]{#2}
\providecommand{\BIBentrySTDinterwordspacing}{\spaceskip=0pt\relax}
\providecommand{\BIBentryALTinterwordstretchfactor}{4}
\providecommand{\BIBentryALTinterwordspacing}{\spaceskip=\fontdimen2\font plus
\BIBentryALTinterwordstretchfactor\fontdimen3\font minus
  \fontdimen4\font\relax}
\providecommand{\BIBforeignlanguage}[2]{{%
\expandafter\ifx\csname l@#1\endcsname\relax
\typeout{** WARNING: IEEEtran.bst: No hyphenation pattern has been}%
\typeout{** loaded for the language `#1'. Using the pattern for}%
\typeout{** the default language instead.}%
\else
\language=\csname l@#1\endcsname
\fi
#2}}
\providecommand{\BIBdecl}{\relax}
\BIBdecl

\bibitem{FranceschettiMinero:ElementsNCS}
M.~Franceschetti and P.~Minero, ``Elements of information theory for networked
  control systems,'' in \emph{Information and Control in Networks}, G.~Como,
  Ed.\hskip 1em plus 0.5em minus 0.4em\relax Springer, 2014, ch.~1, pp. 3--37.

\bibitem{NetworkedControlSurvey_ProcIEEE}
J.~P. Hespanha, P.~Naghshtabrizi, and U.~Xu, ``A survey of recent results in
  networked control systems,'' \emph{Proc.\ IEEE}, vol.~95, no.~1, pp.
  138--162, Jan. 2007.

\bibitem{BansalBasar:JSCC4Control}
R.~Bansal and T.~Ba{\c{s}}ar, ``Simultaneous design of measurement and control
  strategies for stochastic systems with feedback,'' \emph{Automatica},
  vol.~25, no.~5, pp. 679--694, Sep. 1989.

\bibitem{SilvaDerpichOstergaard:ECDQ4Control}
E.~I. Silva, M.~S. Derpich, and J.~{\O}stergaard, ``A framework for control
  system design subject to average data-rate constraints,'' \emph{IEEE Trans.
  Autom.\ Control}, vol.~56, no.~8, pp. 1886--1899, Aug. 2011.

\bibitem{SchenatoSinopoliFranceschettiPoolaSSS}
L.~Schenato, B.~Sinopoli, M.~Franceschetti, K.~Poola, and S.~S. Sastry,
  ``Foundations of control and estimation over lossy networks,'' \emph{Proc.\
  IEEE}, vol.~95, no.~1, pp. 163--187, Jan. 2007.

\bibitem{Nair:L1}
G.~N. Nair, F.~Fagnani, S.~Zampieri, and R.~J. Evans, ``Feedback control under
  data rate constraints: An overview,'' \emph{Proc.\ IEEE}, vol.~95, no.~1, pp.
  108--137, Jan. 2007.

\bibitem{TatikondaSahaiMitter}
S.~Tatikonda, A.~Sahai, and S.~K. Mitter, ``Stochastic linear control over a
  communication channel,'' \emph{IEEE Trans. Autom.\ Control}, vol.~49, no.~9,
  pp. 1549--1561, Sep. 2004.

\bibitem{YukselBasarBook}
S.~{Y\"uksel} and T.~Ba\c{s}ar, \emph{Stochastic Networked Control Systems:
  Stabilization and Optimization Under Information Constraints}.\hskip 1em plus
  0.5em minus 0.4em\relax Boston: Birkh{\"a}user, 2013.

\bibitem{LQGoverAWGN:linear:NoBraslavsky}
J.~S. Freudenberg, R.~H. Middleton, and V.~Solo, ``Stabilization and
  disturbance attenuation over a {Gaussian} communication channel,'' \emph{IEEE
  Trans. Autom.\ Control}, vol.~55, no.~3, pp. 795--799, Mar. 2010.

\bibitem{StavrouSkoglund:ControlWZ:TechRep2019}
P.~A. Stavrou and M.~Skoglund, ``Optimization and tracking of scalar-valued
  {LQG} control under communication link with synchronized or delayed {CaSI} at
  the decoder,'' Division of Information Sciences and Engineering, KTH Royal
  Institute of Technology, Tech. Rep., 2019.

\bibitem{KostinaHassibi:RDF4Control:AC}
V.~Kostina and B.~Hassibi, ``Rate--cost tradeoffs in control,'' \emph{IEEE
  Trans. Autom.\ Control}, Apr., accepted 2019.

\bibitem{JSCC4Control:AC2019}
\khina\CoRes, E.~{Riedel G\r{a}rding}\Student, G.~M. Pettersson\Student,
  V.~Kostina\PI, and B.~Hassibi\PI, ``Control over {Gaussian} channels with and
  without source--channel separation,'' \emph{IEEE Trans. Autom.\ Control},
  vol.~64, no.~9, pp. 3690--3705, Sep. 2019.

\bibitem{JointWZ-WDP}
Y.~Kochman and R.~Zamir, ``Joint {W}yner-{Z}iv/dirty-paper coding by
  modulo-lattice modulation,'' \emph{IEEE Trans.\ Inf.\ Theory}, vol.~55, pp.
  4878--4899, Nov. 2009.

\bibitem{ZamirBookKochmanChapter}
------, \emph{Lattice coding for signals and networks}.\hskip 1em plus 0.5em
  minus 0.4em\relax Cambridge: Cambridge University Press, 2014, ch.
  Modulo-Lattice Modulation.

\bibitem{Tuncel_ZeroDelayJSCCwithWZ}
X.~Chen and E.~Tuncel, ``Zero-delay joint source--channel coding using hybrid
  digital--analog schemes in the {Wyner--Ziv} setting,'' \emph{IEEE Trans.\
  Comm.}, vol.~62, no.~2, pp. 726--735, Feb. 2014.

\bibitem{AastromMurray:book}
K.~J. {\AA}str{\"o}m and R.~M. Murray, \emph{Feedback systems: An Introduction
  for Scientists and Engineers}.\hskip 1em plus 0.5em minus 0.4em\relax
  Princeton university press, 2010.

\bibitem{ElGamalKimBook}
A.~{El Gamal} and Y.-H. Kim, \emph{Network Information Theory}.\hskip 1em plus
  0.5em minus 0.4em\relax Cambridge University Press, 2011.

\bibitem{Elias57:JSCC:BW-expansion}
P.~Elias, ``Channel capacity without coding,'' in \emph{Proc.\ IRE}, vol.~45,
  no.~3, Jan. 1957, pp. 381--381.

\bibitem{WeissmanElGamal_FiniteLookAhead}
T.~Weissman and A.~{El Gamal}, ``Source coding with limited-look-ahead side
  information at the decoder,'' \emph{IEEE Trans.\ Inf.\ Theory}, vol.~52, pp.
  5218--5239, Dec.~2006.

\bibitem{BertsekasControlVol1}
D.~P. Bertsekas, \emph{Dynamic Programming and Optimal Control}, 2nd~ed.\hskip
  1em plus 0.5em minus 0.4em\relax Belmont, MA, USA: Athena Scientific, 2000,
  vol.~I.

\bibitem{ZamirBook}
R.~Zamir, \emph{Lattice Coding for Signals and Networks}.\hskip 1em plus 0.5em
  minus 0.4em\relax Cambridge: Cambridge University Press, 2014.

\end{thebibliography}
\end{document}